\documentclass[aps,pre,twocolumn]{revtex4-2}
\usepackage{graphicx} 
\usepackage{xcolor} 
\usepackage{amsmath,amssymb,bm,amsfonts}
\usepackage{float,wrapfig,tabularx,array}
\usepackage[normalem]{ulem}
\usepackage{mathptmx}

\begin{document}

\title{Phase Separation in a Mixture System of Active and Passive Particles}

\author{Reo Arakaki$^{1,2}$, Tetsuya Hiraiwa$^{*2,3}$, Ryo Akiyama$^{*1}$}

\affiliation{$^{1}$Department of Chemistry, Kyushu University, Fukuoka, Japan\\
$^{2}$Institute of Physics, Academia Sinica, Taipei, Taiwan\\
$^{3}$Physics Division, National Center for Theoretical Sciences, Taipei, Taiwan}

\begin{abstract}
We carried out molecular dynamics simulations of a mixture of self-propelled particles (SPPs) and passive particles. When the self-propulsion speed was sufficiently high, the system separated into two phases. One phase consists of ordered passive particles, and the other is a disordered mixture of both particle types. The separation arises from the interplay between the self-propelled motion of SPPs and thermal motion of the passive particles, since phase separation does not appear with either particle type alone. As the self-propulsion speed increases, the SPPs push surrounding particles away, effectively increasing their excluded size. As a result, the increase in the effective excluded sizes promotes the accumulation of passive particles and the segregation of SPPs into the dilute phase through a mechanism analogous to the depletion interaction. This mechanism explains the phase separation observed in the present study. Our findings provide insights into the potential applications of SPPs and contribute to our understanding of natural systems composed of elements with different levels of motility.
\end{abstract}

\maketitle

\section{Introduction}
Self-propelled particles (SPPs) are particles or, more generally, entities that can convert energy from their environment or internal sources into translational motion \cite{microorganism, microorganism2, SeveralSPPs, nanoMotor}. Collective motion of SPPs is observed in a variety of systems, including schools of fish, clusters of microorganisms, and aggregations of artificial phoretic colloids \cite{SeveralCollectiveMotion, SeveralCollectiveMotion2, CollectiveMotionOfArtificialSPPsByLight}. The behavior of SPP systems cannot be explained within the framework of equilibrium thermodynamics because they are intrinsically out of equilibrium. Studies of such systems have shown that simple interparticle interactions can give rise to striking forms of self-organization. For example, swarms can emerge when particles align their directions of motion with those of their neighbors \cite{21}, and SPPs can undergo motility-induced phase separation (MIPS), in which dense and dilute regions spontaneously form due to persistent self-propulsion \cite{22}.

Previous studies have also investigated how SPPs influence surrounding passive particles. For example, experimental and numerical studies of active–passive mixtures (\cite{23,24} and \cite{25,26}, respectively) have reported intriguing behaviors, including superdiffusion and clustering of passive particles. In the experimental studies, passive colloids were introduced into microorganism suspensions \cite{23,24}, whereas the numerical studies considered mixtures of MIPS-inducing SPPs and large passive particles \cite{25}. Nevertheless, systematic studies of active–passive binary mixtures remain limited. In particular, most previous studies have focused on how the collective motion of SPPs affects passive particles, and the systems investigated have involved complexities, such as biological interaction, collective self-organization of SPPs, anisotropy of particle shape, and size asymmetry. Therefore, the behavior of active–passive mixtures is not yet fully understood, even in simple systems in which all particles are identical disks interacting only through steric repulsion.

When SPPs are introduced into a passive-particle system, two competing effects may arise:
\begin{enumerate}
\item\textit{SPPs promote structural ordering:} The presence of SPPs increases the collision frequency between particles \cite{37}. This may effectively increase the pressure of the system and enhance structural order. 

\item\textit{SPPs induce structural disorder:} SPPs can disrupt the surrounding structure, resulting in a more uniform particle distribution. A previous study reported that highly active SPPs can destroy passive crystalline structures \cite{28}. 
\end{enumerate}
The influence of SPPs on surrounding passive particles is therefore not straightforward. It may depend on several factors, such as system density, composition, and the relative velocity of SPPs compared with passive particles. Moreover, the relationship between self-propulsion and the resulting particle structure remains unclear.

The concept of effective size may provide a link between self-propulsion and the structure of active–passive mixtures. Because the motion of SPPs displaces surrounding particles, their effective excluded area increases. Tanaka et al. reported an effective attraction between two SPPs immersed in a bath of passive particles, which they attributed to a depletion-like mechanism \cite{27}. This discussion draws on an analogy with depletion interactions in equilibrium systems \cite{8',8}. These interactions between larger particles arise from the translational motion of small particles, and the effective interaction is attractive.

We expect that the depletion-like effect may induce phase separation in the SPP–passive particles mixture because, under certain conditions, a binary hard-particle mixture separates into a larger-particle crystal and another phase at equilibrium \cite{3DHSPS1, 3DHSPS2, 3DHSPS3, 3DHSPS4, 3DHSPS5, 3DHSPS6, 3DHSPS7, 3DHSPS8, 3DHSPS9, 3DHSPS10, 3DHSPS11, 3DHSPS12, 3DHSPS13, 3DHSPS14, 3DHSPS15, 3DHSPS16, 36, 2DHDPS2}. This phase separation is induced by effective attraction between the larger particles, which is stronger than the effective attractions between the larger and smaller particles and between the smaller particles. These effective attractions arise from the depletion mechanism.

If the behavior of SPP–passive particle mixtures can be explained in terms of the effective size of SPPs, thermodynamic concepts such as density and depletion effects could provide a framework for describing these systems. Although such an approach may be limited to systems without complex interactions, it could provide a starting point for a unified understanding of the behavior of active–passive mixtures.

The previous study \cite{ActivePassive} observed phase separation in mixtures of SPPs and passive particles. In the model used in Ref.~\cite{ActivePassive}, both SPPs and passive particles are isotropic disks with identical radii, suggesting that the observed phase separation arises solely from activity. However, it remains unclear whether the phase separation originates from collective behavior of the SPPs or from difference in motility between the two components, since the pure SPP system in that model can exhibit MIPS.

In this article, we conducted MD simulations of mixtures of SPPs and passive particles. To clarify the effects of self-propulsion on the structure of the mixtures without the influence of collective behavior, we used an SPP model that exhibits neither swarming nor MIPS in a pure system (Appendix~\ref{Appendix:single component SPPs system}). This approach allowed us to isolate the effects of self-propulsion from collective behavior. We focused on how particle structure depends on the strength of self-propulsion at fluid-like densities. The observed phase separation can be explained in terms of the effective size of SPPs. For comparison, we also performed simulations of equilibrium mixtures of large and small particles (Appendix~\ref{Appendix:binary mixtures of large and small WCA-particles}) to facilitate interpretation of the results.

The remainder of this paper is organized as follows. In Sec.~\ref{Sec:Model}, we describe the simulation model and methods. The results and discussion are presented in Sec.~\ref{Sec:Result and discussion}, beginning with the observed phase separation in Sec.~\ref{Subsec:the phase separation}. We then analyze the phenomenon based on the radial distribution function (Sec.~\ref{Subsec:the radial distribution function}), cluster growth (Sec.~\ref{Subsec:The time evolution of maximum cluster size}), and the kinetic energy of the SPPs (Sec.~\ref{Subsec:the kinetic energy of the SPPs}). Finally, we summarize our findings in Sec.~\ref{Sec:Conclusion}.

\section{Model}\label{Sec:Model}
\subsection{Particle model}
Two-dimensional MD simulations are conducted. The systems consist of SPPs and passive particles, all modeled as disks with the same mass $m$ and diameter $\sigma$. Their positions are updated according to:

\begin{equation}\label{equ:P1}
\frac{d\bm{r}_i}{dt}=\bm{v}_i,
\end{equation}
where $\bm{r}_i$ and $\bm{v}_i$ represent the position and velocity of the $i$-th particle.

\begin{figure}[htbp]
\centering
\includegraphics[width=246pt,clip]{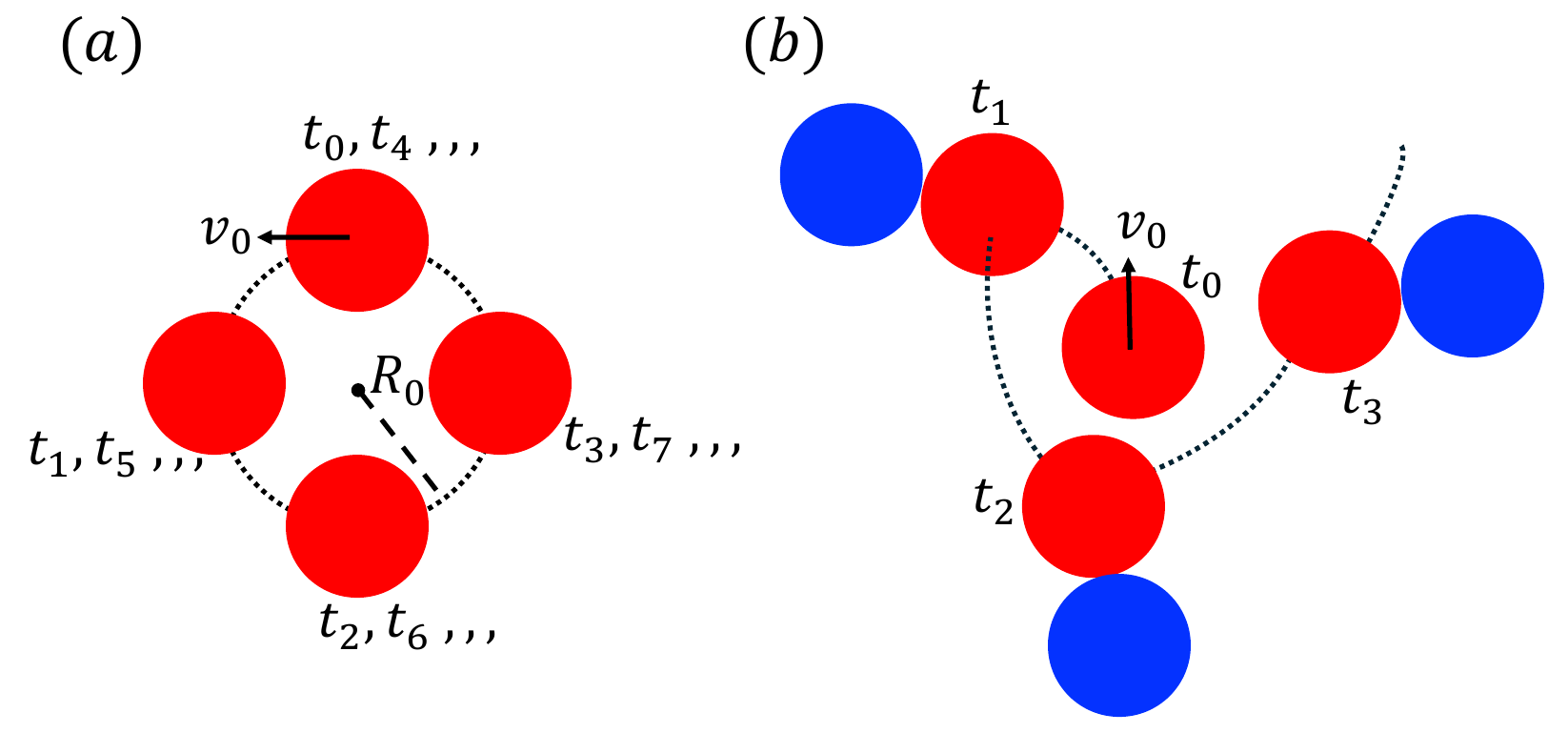}
\caption[]{Schematic illustration of the time evolution of an SPP (a) without collision and (b) with collisions. $v_0$ and $t_0$ represent the initial velocity and time, whereas $t_1,t_2,,,$ denote later times.}
\label{Fig:SPPModel}
\end{figure}

We adopt an inertial orbiting-disk model for the SPPs, with the aim of suppressing MIPS. An isolated SPP (i.e., in the absence of collisions) moves in uniform circular motion with a radius $R_0$ [Fig.~\ref{Fig:SPPModel}(a)]. At the density considered in our simulations, however, the orbital motion of an SPP is difficult to observe because the time between collisions is shorter than that required to complete one rotation [Fig.~\ref{Fig:SPPModel}(b)]. In the absence of passive particles, this SPP model does not exhibit collective behavior within the parameter range investigated in this study (Appendix~\ref{Appendix:single component SPPs system}). The velocity of an SPP obeys:

\begin{equation}\label{equ:SPP1}
\frac{d\bm{v}_i}{dt}=\frac{\bm{f}^{\mathrm{INT}}_i+\bm{f}^{\mathrm{SP}}_i}{m},
\end{equation}
where $\bm{f}^{\mathrm{INT}}_i$ denotes the interaction force exerted on the $i$-th particle. The self-propulsion force is defined as $\bm{f}^{\mathrm{SP}}_i=mv^{\mathrm{SP}}_0\omega\left(-\sin{\theta_i},\cos{\theta_i}\right)$, with $\theta_i=\theta_i^0+\omega t$. Here, $v^{\mathrm{SP}}_0$ and $\omega$ are self-propulsion speed and angular velocity, respectively. The initial value of $\theta_i^0$ is randomly selected from the range $-\pi \leq \theta_i < \pi$ in the initial state.

\begin{figure*}[t]
\centering
\includegraphics[width=\textwidth,clip]{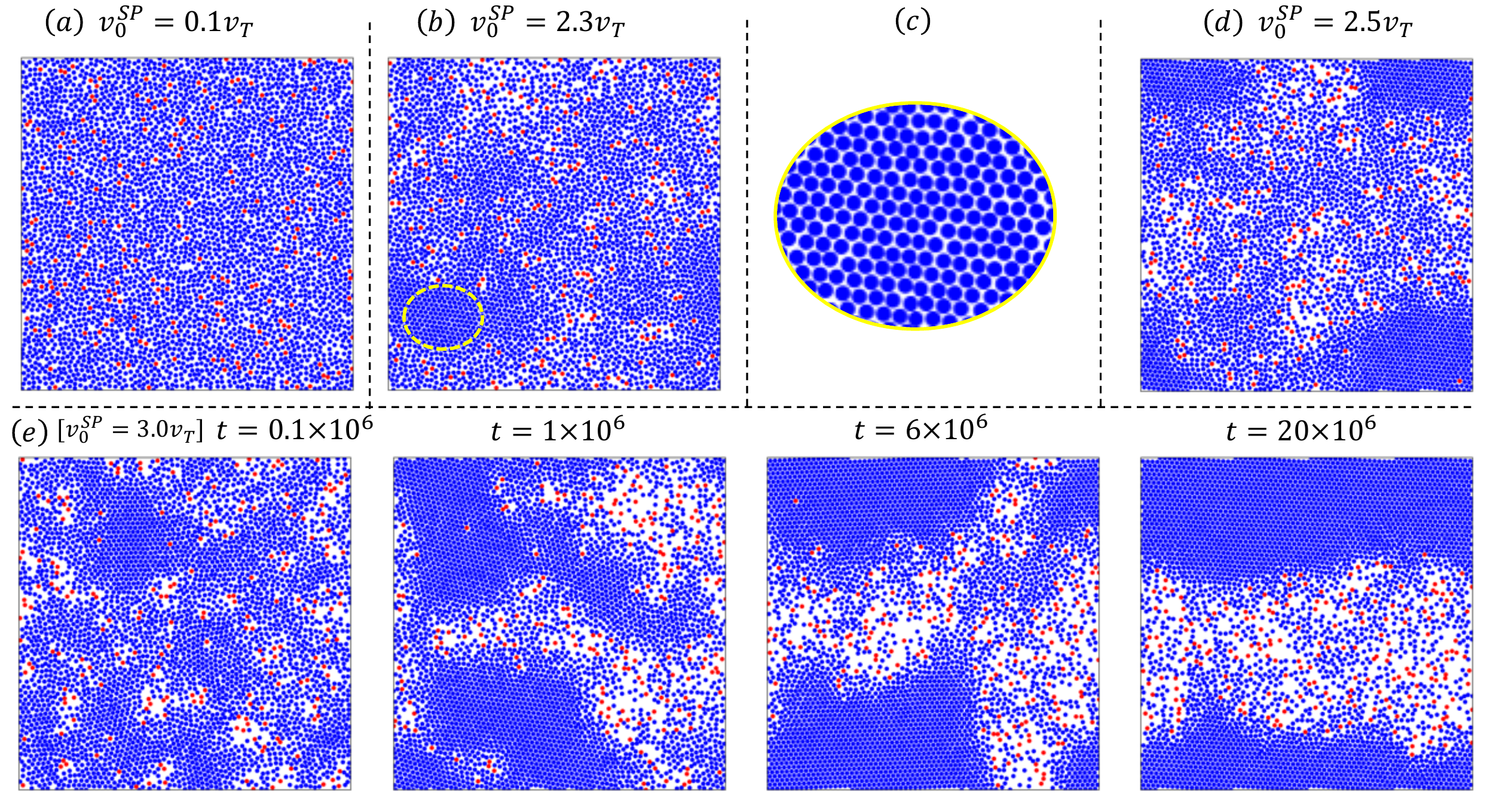}
\caption[]{Snapshots of the simulations: (a) $v_0^{\mathrm{SP}}=0.1v_T$, (b) $v_0^{\mathrm{SP}}=2.3v_T$, (c) an enlarged view of the yellow circle in (b), (d) $v_0^{\mathrm{SP}}=2.5v_T$, and (e) $v_0^{\mathrm{SP}}=3.0v_T$ at $t=(0.1, 1, 6, 20)\times10^6$.  The red and blue disks represent SPPs and passive particles, respectively.}
\label{Fig:snapshots}    
\end{figure*}

Passive particles do not have a self-propulsion force. We assume that they are coupled to a heat bath. In this study, we maintain the kinetic energy of passive particles to reproduce heat-bath-like behavior. In systems containing SPPs, the Langevin approach fails to accurately control the kinetic energy of passive particles, with its average deviating from the target value. To overcome this, we adopt the Nosé–Hoover method. The velocity of a passive particle obeys:

\begin{equation}\label{equ:PP1}
\frac{d\bm{v}_i}{dt}=\frac{\bm{f}^{\mathrm{INT}}_i}{m}-\zeta\bm{v}_i,
\end{equation}
\begin{equation}\label{equ:PP2}
\frac{d\zeta}{dt}=\frac{1}{Q}\left(\sum_{j=1}^{N_P}{m\bm{v}_j^2}-g\mathrm{k_B}T\right).
\end{equation}
Here, the thermostat variable $\zeta$ acts as a friction coefficient, and the parameter $Q$ represents the thermostat mass. $N_P$ denotes the number of passive particles, and $g$ represents the degrees of freedom, defined as $g\equiv2N_P+1$. Finally, $\mathrm{k_B}$ is the Boltzmann constant and $T$ denotes the target temperature.

Equation~\ref{equ:PP2} describes the time evolution of the friction coefficient. When the kinetic energy of the passive particles exceeds the target value, the friction increases, slowing down the particles. Conversely, when the kinetic energy falls below the target value, the friction decreases and may even become negative, thereby accelerating the particles. As a result, the kinetic energy of passive particles fluctuates around the target energy. Because we assume that the number of passive particles is sufficiently large relative to the number of SPPs, the system temperature fluctuates around the target value according to Eqs. ~\ref{equ:P1}, \ref{equ:PP1}, and \ref{equ:PP2}, provided that the activity of the SPPs is not too high.

In this study, we adopt the Weeks–Chandler–Andersen (WCA) potential as the interparticle interaction potential. The WCA potential incorporates only the repulsive term arising from excluded-volume effects. This potential is given by:
\begin{equation}\label{equ:WCA}
V_{ij}\left(r\right)=4\epsilon\left[\left(\frac{\sigma}{r}\right)^{12}-\left(\frac{\sigma}{r}\right)^{6}\right]+\epsilon,
\end{equation}
if $r\le 2^{1/6}\sigma$, and $V_{ij}\left(r\right) =0$ otherwise, where $r=\left|\bm{r}_i-\bm{r}_j\right|$. The $\epsilon$ and $\sigma$ represent the energy scale and the reference diameter, respectively. The interaction force exerted on the $i$-th particle is calculated as $\bm{f}^{\mathrm{INT}}_i=-\frac{\partial}{\partial \bm{r}_i}\sum{V_{ij}}$.

\subsection{System setting}
The simulation box is square, $L\times L$, and surrounded by periodic boundary conditions. The system consists of SPPs and passive particles. Among all particles, 5$\%$ are SPPs ($N_S=200$) and 95$\%$ are passive particles ($N_P=3800$).

The angular velocity $\omega$ is set to $v_0^{\mathrm{SP}}/R_0$, so that a single free SPP moves in a uniform circular motion of radius $R_0$. The $R_0$ is set to $\sigma$. In our simulations, the main parameter is $v_0^{\mathrm{SP}}$. We vary $v_0^{\mathrm{SP}}$ from $0.1v_T$ to $3.0v_T$. Here, $v_T$ denotes the “thermal speed”, defined by the two-dimensional kinetic energy relation $\frac{1}{2} mv_T^2=\mathrm{k_B}T$.

The units of length, mass, and energy are set to $\sigma$, $m$, and $\epsilon$, respectively. The unit of time is also set as $\tau=\sqrt{m\sigma^2/\epsilon}$. The temperature $T$ and the number density $(N_S+N_P)/L^2$ are set to $0.722 \epsilon/\mathrm{k_B}$ and $0.700\sigma^{-2}$, respectively, so that the system behaves in a liquid-like manner when the self-propulsion of SPPs is zero. The time step $\Delta t$ is set as $9.30\times10^{-4} \tau$. Each simulation is run until $2\times10^7\Delta t$. The thermostat mass $Q$ is set as $5.0m\sigma^2$. With this value, the total kinetic energy shows no unusual behavior over time.

Initially, all particles are randomly positioned without overlapping, with the particles arranged sufficiently close to accommodate all particles in the simulation box. The velocities of passive particles are randomly assigned according to the Maxwell–Boltzmann distribution corresponding to the target temperature $T$. The velocity of each SPP is assigned as a vector with magnitude $v_0^{\mathrm{SP}}$ and a randomly chosen direction.

\section{Result and discussion}\label{Sec:Result and discussion}
\subsection{Phase separation behavior}\label{Subsec:the phase separation}
We performed MD simulations of mixtures of SPPs and passive particles while varying the self-propulsion speed $v_0^{\mathrm{SP}}$ to investigate how the system responds to self-propulsion. SPPs and passive particles separate as the self-propulsion speed increases. Figure~\ref{Fig:snapshots} shows representative snapshots from our simulations. The red and blue disks represent SPPs and passive particles, respectively. At low self-propulsion speeds, $v_0^{\mathrm{SP}}=0.1v_T$ [Fig.~\ref{Fig:snapshots}(a)], the SPPs and passive particles are well mixed. In contrast, some crystal-like clusters composed of passive particles are observed when $v_0^{\mathrm{SP}}=2.3v_T$ [Fig.~\ref{Fig:snapshots}(b-c)]. These clusters repeatedly grow and collapse, with new clusters subsequently forming at different locations. The system seems to be in a steady state. When $v_0^{\mathrm{SP}}\ge 2.5v_T$ [Fig.~\ref{Fig:snapshots}(d) and (e), a large cluster of passive particles forms. The system undergoes macroscopic phase separation into dilute and dense regions. The dilute phase contains both SPPs and passive particles, while the dense phase consists of passive particles. When $v_0^{\mathrm{SP}}\ge2.7v_T$ [Fig.~\ref{Fig:snapshots}(e)], the cluster extends beyond one side of the simulation box.

\begin{figure}[htbp]
\centering
\includegraphics[width=246pt,clip]{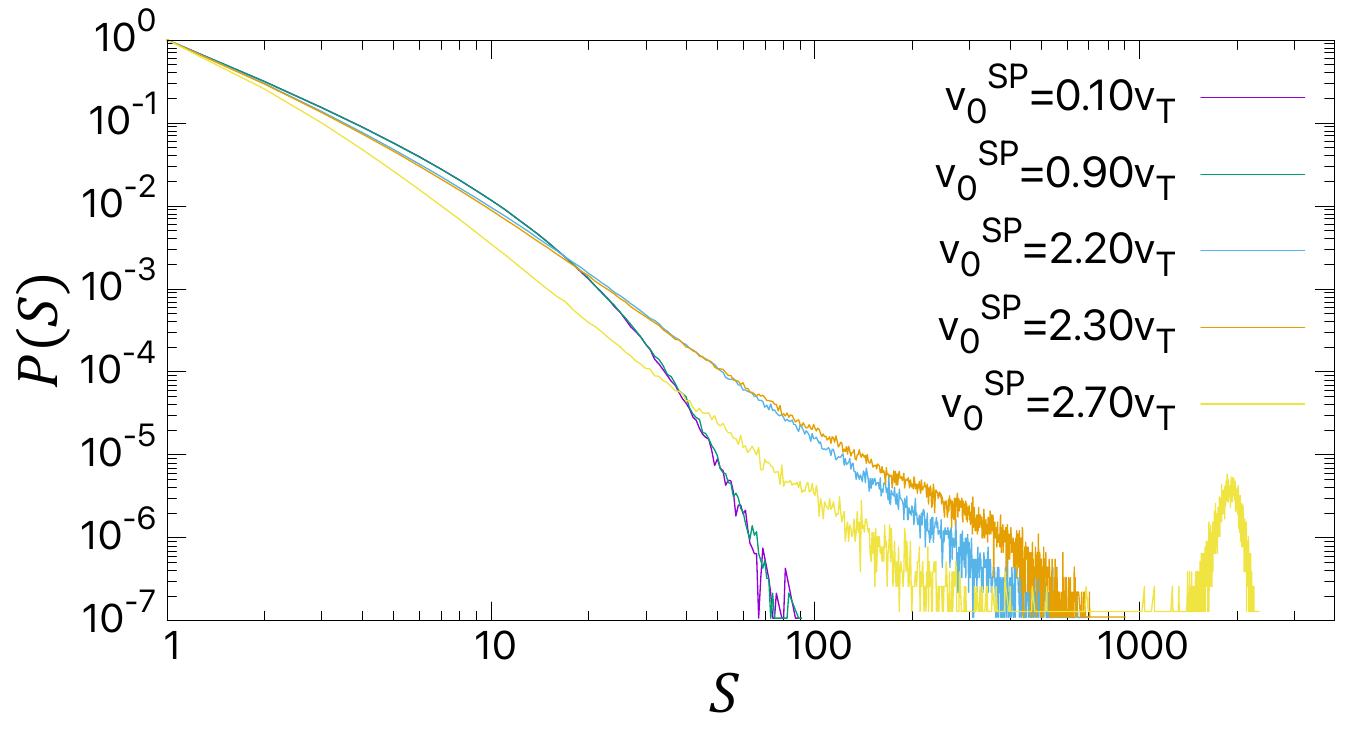}
\caption[]{Cluster size, $S$, distributions. The horizontal axis represents the cluster size. The vertical axis represents the number of clusters scaled by the number of single-particle clusters. Purple; $v_0^{\mathrm{SP}}=0.1v_T$, green; $v_0^{\mathrm{SP}}=0.9v_T$, cyan; $v_0^{\mathrm{SP}}=2.2v_T$, orange; $v_0^{\mathrm{SP}}=2.3v_T$, yellow; $v_0^{\mathrm{SP}}=2.7v_T$.}
\label{Fig:CSD}
\end{figure}

In the present simulation, the transition to the ordered phase occurs in the range of $v_0^{\mathrm{SP}}=2.2v_T$ to $v_0^{\mathrm{SP}}=2.3v_T$. Here, cluster size analysis was performed to estimate transition point. The cluster is defined using a distance-based criterion. When the distance between two particles is less than $1.1\sigma$, they belong to the same cluster. The cluster size is defined as the number of particles in the cluster. This definition has been adopted in various studies of cluster analysis \cite{25}. Figure~\ref{Fig:CSD} shows the cluster size distributions (CSDs), calculated from snapshots taken after $t=10^7 \Delta t$. Here, we define the steady state as $t\ge 10^7\Delta t$, as discussed in Sec.~\ref{Subsec:The time evolution of maximum cluster size}. When $v_0^{\mathrm{SP}}=0.1v_T$, the CSD decays exponentially. This behavior is consistent with clusters formed by random aggregation in disordered states. When $v_0^{\mathrm{SP}}$ exceeds $0.9v_T$, the distribution changes. At $v_0^{\mathrm{SP}}=2.2v_T$, the CSD decays approximately as a power law, although the decay is slightly faster than the power law. At $v_0^{\mathrm{SP}}=2.3v_T$, the decay becomes slower than the power law. These results indicate that the transition from the disordered phase to the ordered phase, in which cluster formation is favored, occurs at $v_0^{\mathrm{SP}}=2.2v_T$-$2.3v_T$. For $v_0^{\mathrm{SP}}=2.7v_T$, a distinct peak is observed at a larger cluster size. In this high self-propulsion speed regime, the CSD can no longer be described by a simple exponential or power-law form.

\begin{figure}[htbp]
\centering
\includegraphics[width=246pt,clip]{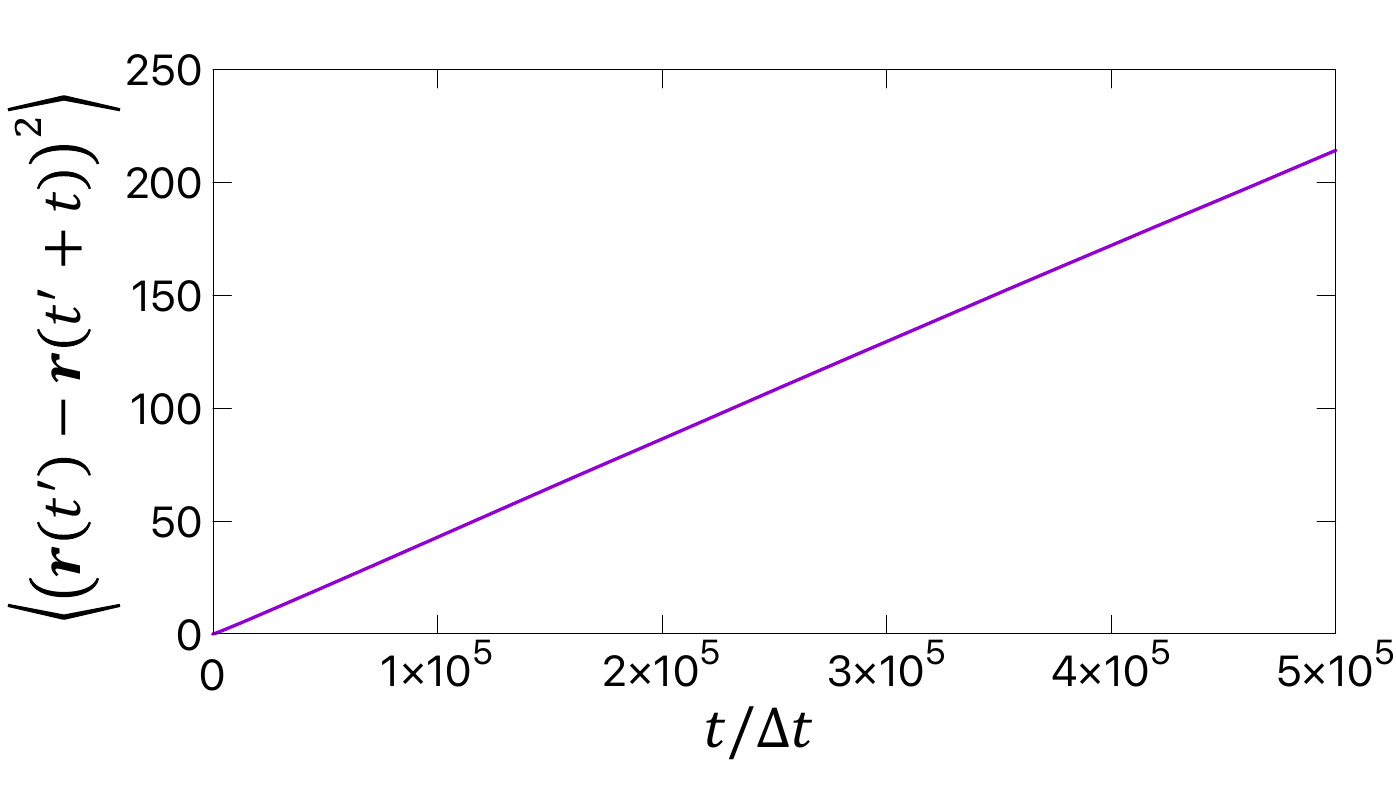}
\caption[]{Mean square displacement (MSD) of the passive particles in a pure system.}
\label{Fig:MSD}
\end{figure}

In the present study, the phase separation arises from the interplay between the SPPs and the passive particles. This phase separation does not arise from the self-organization of SPPs because a system consisting solely of the SPPs used in this study does not exhibit collective self-organization, such as MIPS \cite{22,25} or swarming \cite{21}. Furthermore, the SPPs and passive particles have the same diameter; thus, so-called active depletion \cite{29,30,31} cannot account for the phase separation observed in the present study. Activity-induced crystallization of a glassy phase \cite{28} also cannot explain the clustering in the present simulation because the pure passive-particle system does not exhibit glassy dynamics at the present packing fraction. To confirm this, we show the mean square displacement (MSD) of the pure passive-particle system, which exhibits linear growth at long times [Fig.~\ref{Fig:MSD}], indicating diffusive rather than glassy dynamics.

\subsection{The radial distribution function}\label{Subsec:the radial distribution function}
The radial distribution function (RDF), $g\left(r\right)$, was calculated using snapshots taken at $t>10^7\Delta t$. The RDF is defined as:
\begin{equation}\label{equ:RDF}
g\left(r\right)\equiv\frac{\langle n_i\left(r\right)\rangle}{2\pi r\Delta r}\frac{L^2}{N_{j}},
\end{equation}
where $i$ and $j$ denote the particle types, $i,j \in \{\mathrm{SPP},\mathrm{passive\:particle}\}$. $\langle \cdot \rangle$ denotes a sample average, and $n_i\left(r\right)$ represents the number of $i$-type particles located within the distance between $r$ and $r+\Delta r$ from one $j$-type particle. $N_{j}$ is the total number of $j$-type particles.

\begin{figure}[htbp]
\centering
\includegraphics[width=246pt,clip]{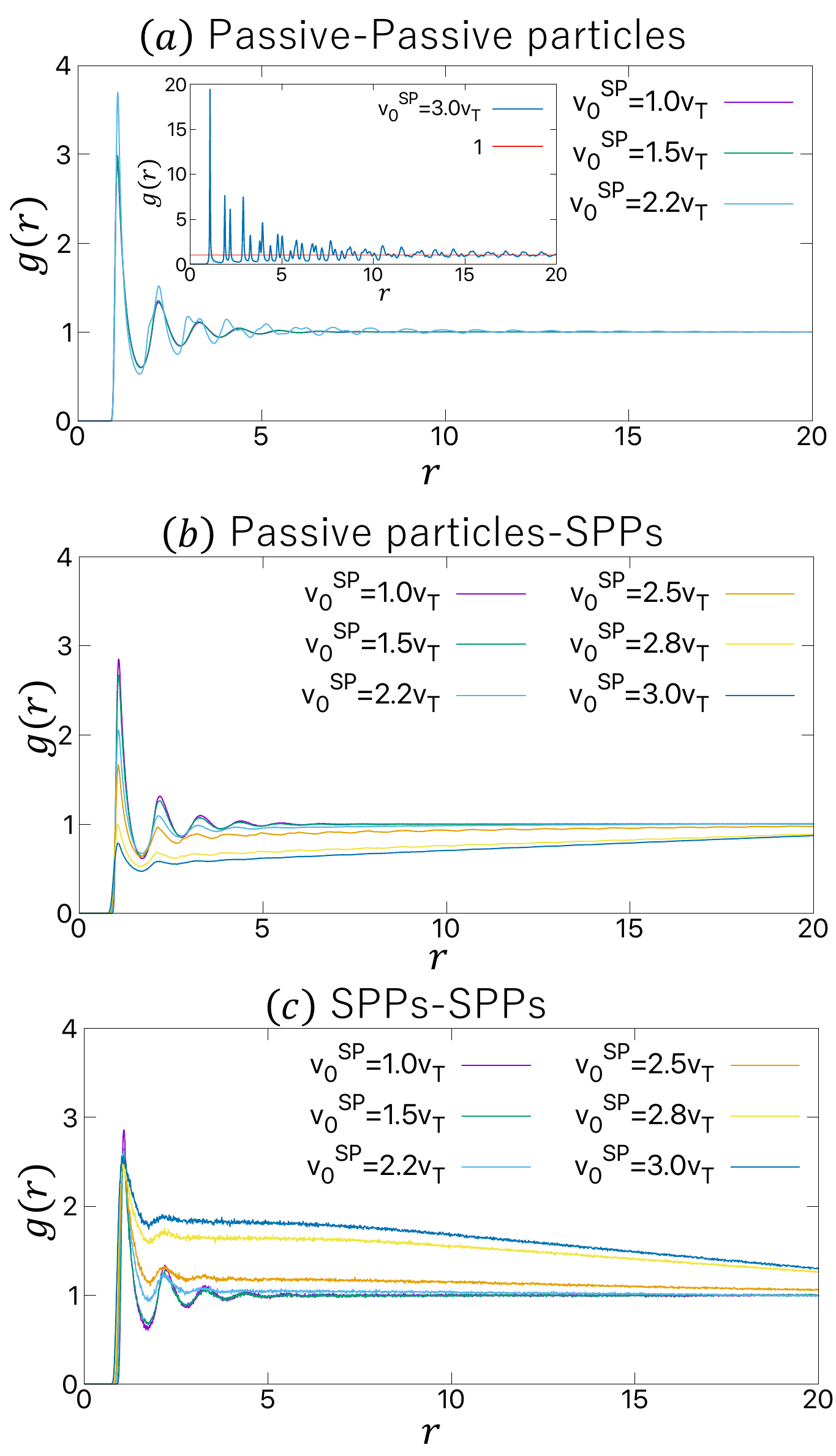}
\caption[]{Radial distribution functions (RDFs) $g\left(r\right)$. (a) $g\left(r\right)$ between passive particles, (b) $g\left(r\right)$ between passive particles and SPPs, and (c) $g\left(r\right)$ between SPPs. Purple; $v_0^{\mathrm{SP}}=1.0v_T$, green; $v_0^{\mathrm{SP}}=1.5v_T$, cyan; $v_0^{\mathrm{SP}}=2.2v_T$, orange; $v_0^{\mathrm{SP}}=2.5v_T$, yellow; $v_0^{\mathrm{SP}}=2.8v_T$, blue; $v_0^{\mathrm{SP}}=3.0v_T$.}
\label{Fig:RDF}
\end{figure}

As the self-propulsion speed increases, the RDF reveals an increasingly ordered arrangement of passive particles. Figure~\ref{Fig:RDF}(a) shows the RDF between passive particles. When the self-propulsion speed is slow ($v_0^{\mathrm{SP}}=1.0v_T$ and $1.5v_T$), the RDFs are fluid-like. The RDFs have a first peak at $r=\sigma$ and exhibit damped oscillations. When the self-propulsion speed increases ($v_0^{\mathrm{SP}}=2.2v_T$), the characteristic peaks associated with crystalline ordering appear in the RDF. When the self-propulsion speed becomes fast ($v_0^{\mathrm{SP}}=3.0v_T$; see inset), these characteristic peaks become significant, suggesting that the solid-like ordered structure has grown in the system.

Figure~\ref{Fig:RDF}(b) shows the RDF between passive particles and SPPs. The height of the first peak decreases with increasing self-propulsion speed. Here, the first peak is defined as the first local maximum of $g\left(r\right)$ encountered when increasing $r$ from $r=0$, and its height is denoted by $g_{\mathrm{peak}}$ below. This dependence on self-propulsion speed indicates a reduction in the number of passive particle–SPP pairs at short distances. Furthermore, when $v_0^{\mathrm{SP}} \ge 2.5v_T$, the RDFs become below $1$ in the range $2<r<20$ and exhibit a long-range increase. This increase in the RDFs indicates that the effective repulsion is long-ranged. As shown in the snapshots [Fig.~\ref{Fig:snapshots}], phase separation becomes more pronounced as the self-propulsion speed increases, consistent with the RDF results.

Figure~\ref{Fig:RDF}(c) shows the SPP–SPP RDF. We now examine the dependence of the first peak on the self-propulsion speed. One would expect this dependence to be similar to that of the passive–passive particle RDF. However, even though the particles are the same, the first peak becomes lower. This could be misinterpreted as behavior contrary to the phase separation observed in the snapshots. However, examination of the long-range behavior reveals that the RDF results are consistent with the observed phase separation. Let us examine the RDF in the range $2<r<20$. When the self-propulsion speed exceeds $2.5v_T$, the value of the RDF in the range exceeds unity and exhibits a long monotonically decreasing profile. This indicates the presence of a long-range effective attraction. This long-range effective attraction intensifies as self-propulsion speed increases.

\begin{figure}[htbp]
\centering
\includegraphics[width=246pt,clip]{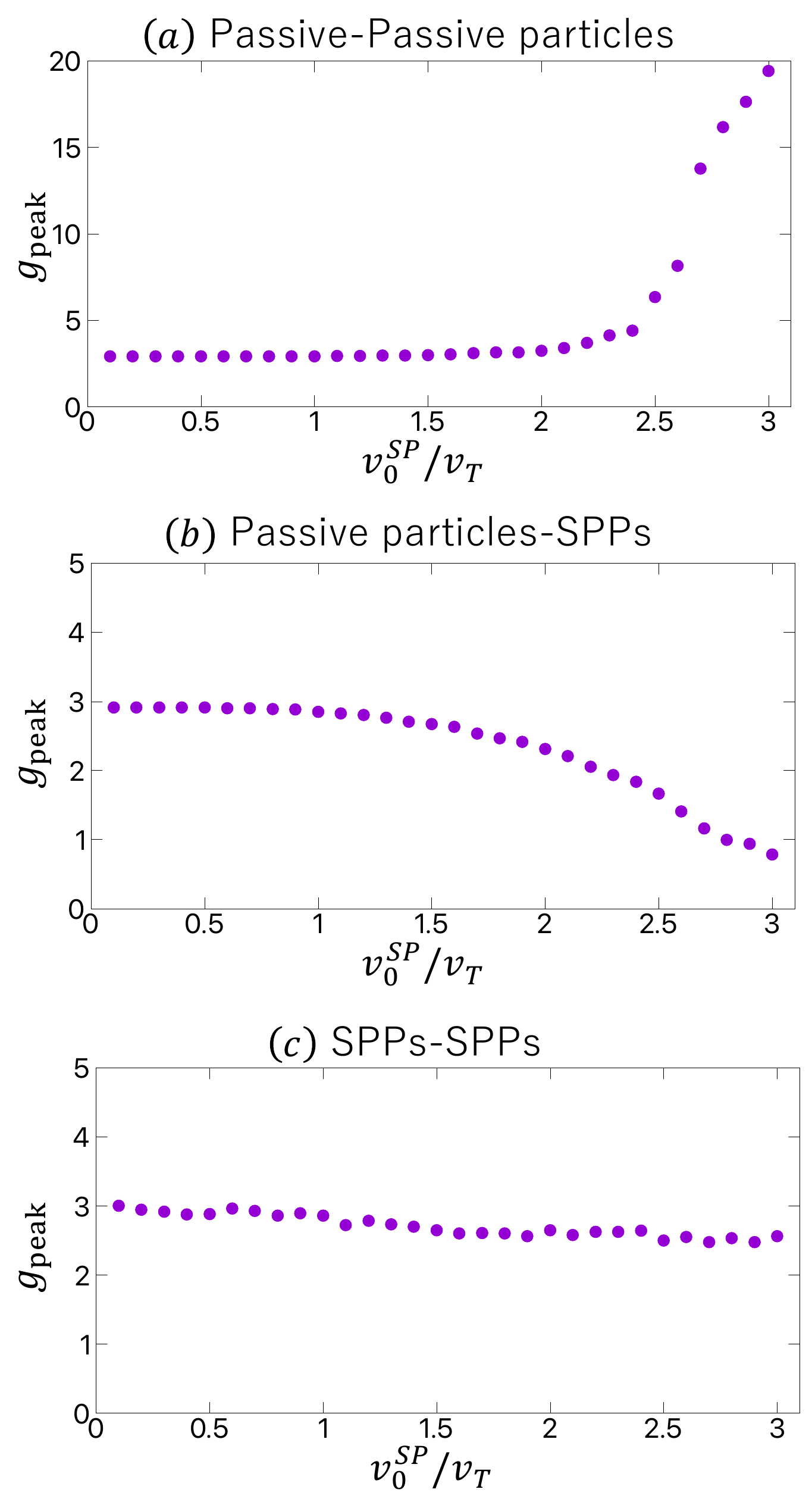}
\caption[]{The height of the first peaks of the radial distribution functions (RDFs) $g_{\mathrm{peak}}$. (a) $g_{\mathrm{peak}}$ between passive particles, (b) $g_{\mathrm{peak}}$ between passive particles and SPPs, and (c) $g_{\mathrm{peak}}$ between SPPs. The jump between $v_0^{\mathrm{SP}}=2.6v_T$ and $2.7v_T$ in panel (a) comes from the finite-size effect.}
\label{Fig:FirstPeakRDF}
\end{figure}

The first peak of the RDF between passive particles increases suddenly when the self-propulsion speed $v_0^{\mathrm{SP}}$ exceeds a certain value. This dependence suggests that the formation of ordered structures does not proceed gradually as $v_0^{\mathrm{SP}}$ increases. To confirm this behavior, it is necessary to plot the $v_0^{\mathrm{SP}}$ dependence of the first peak [Fig.~\ref{Fig:FirstPeakRDF}(a)]. The $g_{\mathrm{peak}}$ remains almost constant up to $v_0^{\mathrm{SP}}=2.1v_T$. However, it increases sharply around $v_0^{\mathrm{SP}}=2.2v_T$-$2.5v_T$, indicating the onset of ordered structures. This region of drastic increase is consistent with the transition point ($v_0^{\mathrm{SP}}=2.2v_T$-$2.3v_T$) estimated from the CSD analysis (Sec.~\ref{Subsec:the phase separation}).

Unlike the first peak for passive particles [Fig.~\ref{Fig:FirstPeakRDF}(a)], the first peak $g_{\mathrm{peak}}$ between passive particles and SPPs [Fig.~\ref{Fig:FirstPeakRDF}(b)] and between SPPs [Fig.~\ref{Fig:FirstPeakRDF}(c)] change gradually with increasing $v_0^{\mathrm{SP}}$. The first peak between the SPPs shows almost no dependence on $v_0^{\mathrm{SP}}$. Although two SPPs move apart following a direct collision, their self-propelled motion causes them to approach each other again. Consequently, the probability of finding two SPPs near one another increases, giving rise to an effective attractive force at $r>2$.

These observations presented in Figs.~\ref{Fig:snapshots}, \ref{Fig:RDF}, and \ref{Fig:FirstPeakRDF} suggest the following “depletion” mechanism. Because SPPs push surrounding passive particles away, they are surrounded by low-density regions, making their effective size larger. If we replaced the effectively large particles with actual large particles, we could explain this separation in terms of the “depletion” mechanism. As the effective size of the SPP increases, the attraction between the SPPs increases. Consequently, passive particles are excluded from the regions surrounding the SPPs, eventually leading to phase separation. This depletion mechanism has been discussed in the context of phase separation of binary mixtures \cite{8', 8, 36, 3DHSPS1, 3DHSPS2, 3DHSPS3, 3DHSPS4, 3DHSPS5, 3DHSPS6, 3DHSPS7, 3DHSPS8, 3DHSPS9, 3DHSPS10, 3DHSPS11, 3DHSPS12, 3DHSPS13, 3DHSPS14, 3DHSPS15, 3DHSPS16, 2DHDPS1, 2DHDPS3}.

However, the present mixture differs from binary hard-particle mixtures. In those cases, the larger particles form an ordered phase. In contrast, the SPPs do not order in our simulations, although their effective size is larger than that of the passive particles. This is because SPPs can penetrate the effective excluded regions of other SPPs. This behavior is reflected in the SPP–SPP RDF [Fig.~\ref{Fig:RDF}(c)]. When the passive particles exhibit ordering ($v_0^{SP} \ge 2.5v_T$), the SPP–SPP RDF is much larger than $1$ within the range of $1 < r < 20$, and the peaks beyond the second peak become less distinct. Self-propulsion drives SPPs toward one another while maintaining varying separation distances, thereby preventing ordering. The expanded effective size of SPPs compresses passive particles into a smaller region, resulting in the ordering of “smaller” passive particles.  

\subsection{The time evolution of maximum cluster size} \label{Subsec:The time evolution of maximum cluster size}
\begin{figure}[htbp]
\centering
\includegraphics[width=246pt,clip]{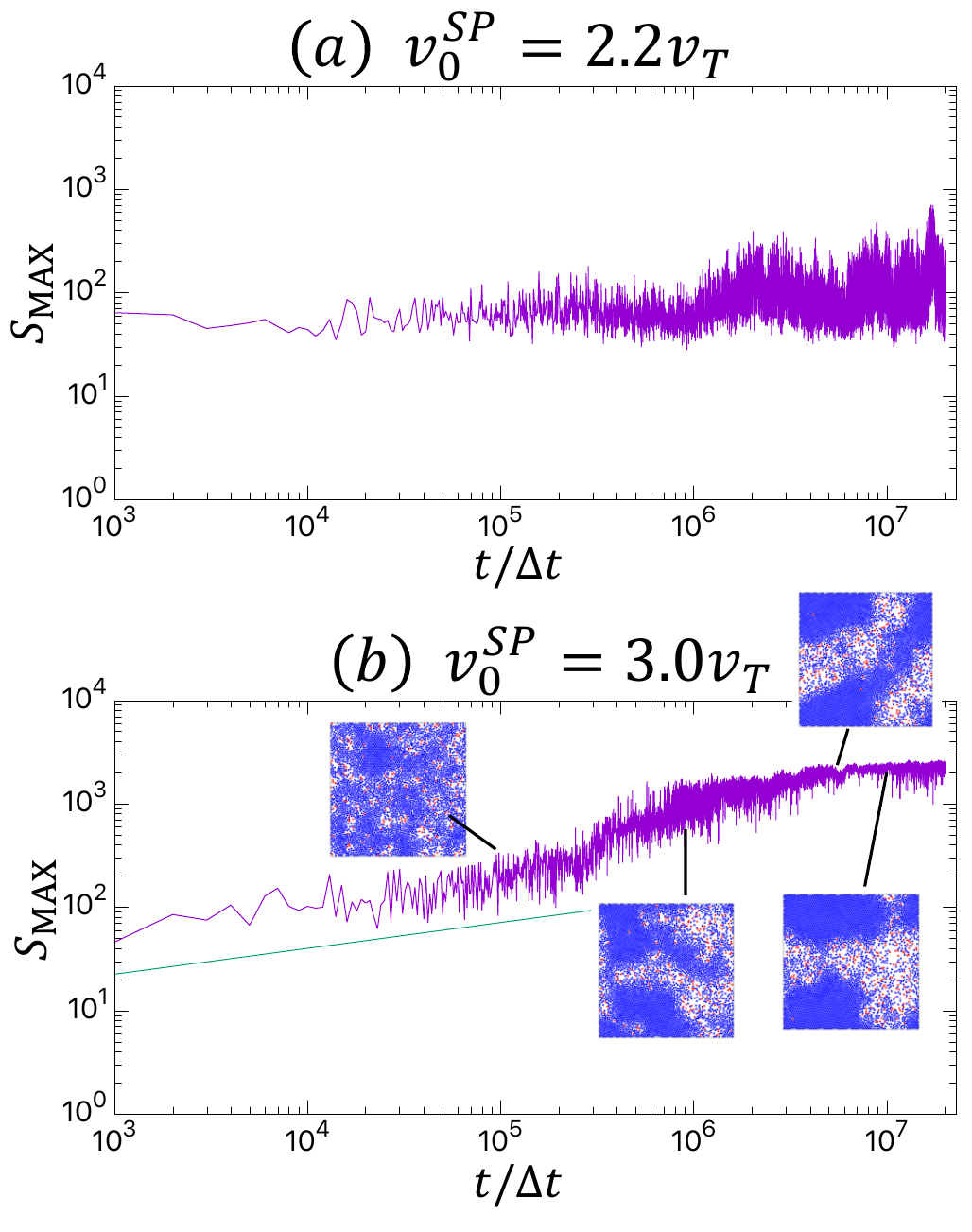}
\caption[]{The time evolution of the maximum cluster size, for (a) $v_0^{\mathrm{SP}}=2.2v_T$, (b) $v_0^{\mathrm{SP}}=3.0v_T$. Green lines show $\propto t^{0.25}$.}
\label{Fig:MaxCluster}
\end{figure}

Figure~\ref{Fig:MaxCluster} shows the time evolution of the maximum cluster size $S_{\mathrm{MAX}}$ (i.e., the size of the largest cluster in the system). Here, the cluster size is defined using the same method described in Sec.~\ref{Subsec:the phase separation}. When $v_0^{\mathrm{SP}}=2.2v_T$ [Fig.~\ref{Fig:MaxCluster}(a)], the maximum cluster size fluctuates around a certain number. Therefore, the system is in a steady state where small clusters appear but do not grow, and macroscopic phase separation does not occur. On the other hand, when $v_0^{\mathrm{SP}}=3.0v_T$ [Fig.~\ref{Fig:MaxCluster}(b)], the maximum cluster size increases and eventually saturates. For all self-propulsion speeds investigated in this study, the maximum cluster size fluctuates around a constant value for $t>10^7\Delta t$. We therefore define the region of $t>10^7\Delta t$ as the steady-state regime.

When the system phase separates, the maximum cluster size increases according to a power law. At $v_0^{\mathrm{SP}}=3.0v_T$ [Fig.~\ref{Fig:MaxCluster}(b)], the maximum cluster size increases with time and eventually saturates at a certain value. This saturation is caused by a finite-size effect of the simulation box. When the gap between the ends of the clusters becomes small, the connection between clusters in neighboring periodic images becomes more likely than further cluster growth. In fact, $S_{\mathrm{MAX}}$ grows rapidly at $t\sim3\times10^5$. Before this rapid growth, the largest cluster grows according to a power law.

\subsection{The time evolution of the kinetic energy of the SPPs} \label{Subsec:the kinetic energy of the SPPs}
\begin{figure}[htbp]
\centering
\includegraphics[width=246pt,clip]{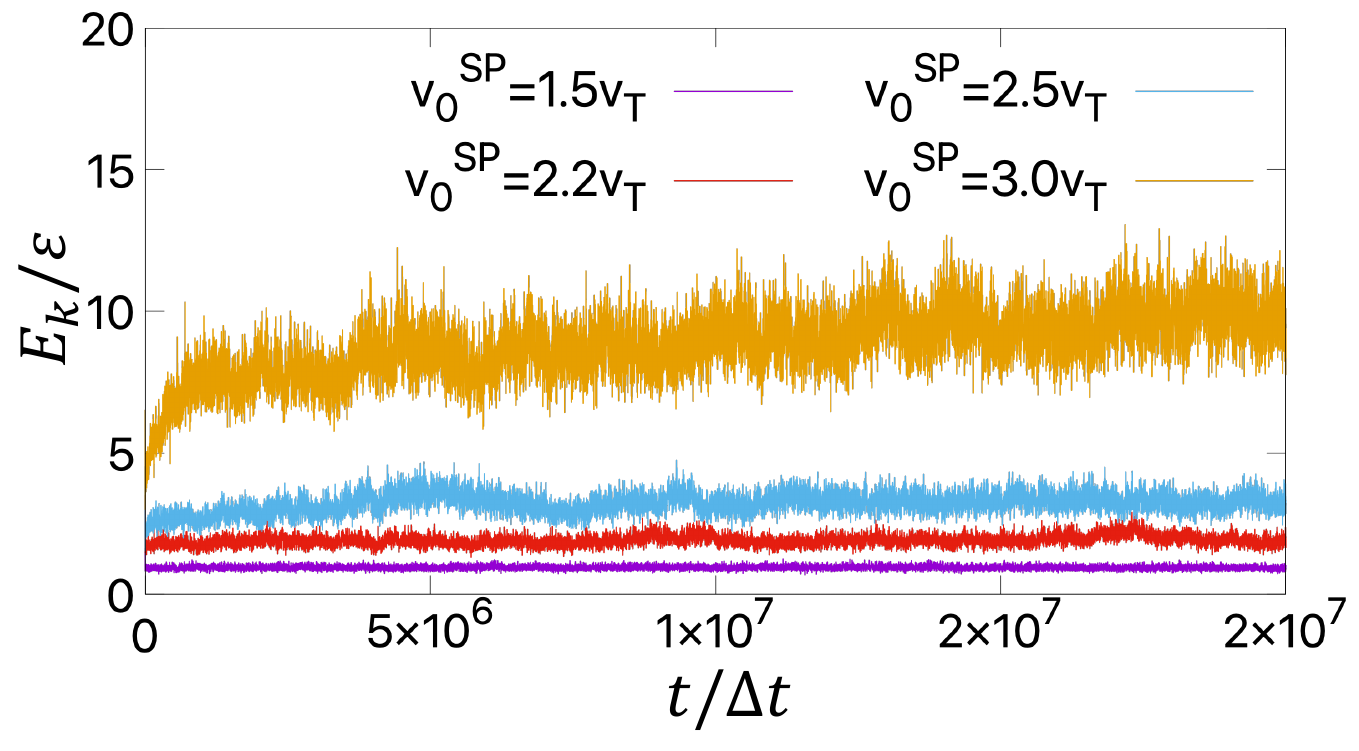}
\caption[]{The time evolution of the kinetic energy of SPPs. Purple; $v_0^{\mathrm{SP}}=1.5v_T$, green; $v_0^{\mathrm{SP}}=2.2v_T$, cyan; $v_0^{\mathrm{SP}}=2.5v_T$, orange; $v_0^{\mathrm{SP}}=3.0v_T$.}
\label{Fig:KineticEnergy_SPP}
\end{figure}

The time evolution of the kinetic energy of SPPs depends on phase separation behavior. Figure~\ref{Fig:KineticEnergy_SPP} shows the time evolution of the kinetic energy, defined by $E_k\equiv\sum_{i=1}^{N_S}m\bm{v}_i^2/2{N_S}$. When the self-propulsion speed is slow ($v_0^{\mathrm{SP}}=1.5v_T$ and $2.2v_T$), the kinetic energy of SPPs fluctuates around a constant value. On the other hand, when the self-propulsion speed is fast ($v_0^{\mathrm{SP}}=2.5v_T$ and $3.0v_T$), the kinetic energy initially increases and then saturates. We can explain this behavior as follows. The time evolution of the kinetic energy depends on the extent to which phase separation develops. As phase separation progresses, the area available for SPP movement increases, leading to an increase in net velocity. Phase separation therefore results in a longer relaxation time than in a system in which phase separation is suppressed. The dependence of the average kinetic energy of SPPs in the steady state on the self-propulsion speed is discussed in Appendix~\ref{Appendix:The average kinetic energy of SPPs}.

\section{Conclusion}\label{Sec:Conclusion}
Our simulations demonstrate that motility heterogeneity can induce phase separation in a mixture of SPPs and passive particles. As the self-propulsion speed increases, the system transitions from a mixed state to a phase-separated state consisting of a passive particle-rich dense region and a mixed dilute region. This phase separation is not driven by SPP ordering, swarming, or other collective motions of SPPs because the pure SPP system does not exhibit swarming or MIPS under the present conditions. Rather, the phase separation arises from the interaction between SPPs and passive particles. In particular, an increase in self-propulsion speed effectively expands the excluded size of the SPPs, inducing effective attraction between SPPs through the depletion-like mechanism, and restricting the space available to passive particles.

Phase separation in the present system appears analogous to that observed in binary hard-disk systems, which can be explained by the depletion effect. Because SPPs collide with other particles and transfer substantial momentum to them, the particle density decreases in the vicinity of SPPs. Thus, the excluded area of the SPP is effectively expanded in the system. The increase in the effective excluded area becomes larger as the self-propulsion speed increases. Therefore, the present system resembles a two-component hard-disk system with respect to the exclusion effect.

However, according to literature \cite{3DHSPS1, 3DHSPS2, 3DHSPS3, 3DHSPS4, 3DHSPS5, 3DHSPS6, 3DHSPS7, 3DHSPS8, 3DHSPS9, 3DHSPS10, 3DHSPS11, 3DHSPS12, 3DHSPS13, 3DHSPS14, 3DHSPS15, 3DHSPS16, 36, 2DHDPS2}, the large particles order in a two-component hard-particle system. In contrast, the passive particles, which are effectively smaller, form an ordered phase in the present simulation. This difference can be explained in terms of the SPP–SPP RDF. In the case of the hard-particle mixture, other large particles cannot penetrate the excluded area surrounding a large particle. Thus, the contact distance is fixed, and the highest peak of the RDF clearly exists at the contact distance. This is because a large particle maintains an approximately fixed distance from its neighboring particles due to excluded-volume effects. Consequently, the large particles can form an ordered structure.

In contrast, an SPP collides with other particles, but other SPPs can exist at various distances from the SPP owing to the self-propulsion motion. As a result, the contact distance becomes unclear, and the RDF remains larger than $1$ for $1<r<20$. The RDF indicates that SPPs cannot maintain a fixed distance from their neighboring particles because of their self-propelled motion. Therefore, the SPPs cannot form an ordered structure. The passive particles are compressed into a smaller area, and the packing fraction increases due to the expansion of the SPPs’ effective area. Because passive particles maintain approximately fixed distances from their neighboring particles due to repulsive-core interactions, they can form an ordered structure.

We can also compare the phase separation in the present study with that in other models of SPP–passive particle mixtures. The phase separation observed in our simulations differs from that in Ref.~\cite{ActivePassive}. In Ref.~\cite{ActivePassive}, mixtures of SPPs and passive particles separate into dense and dilute phases, and then the SPPs and passive particles further segregate within the dense phase. In contrast, in the present system, the dense phase consists only of passive particles, and the dilute phase consists of both SPPs and passive particles. This difference can be understood in terms of MIPS. In the model used in Ref.~\cite{ActivePassive}, the SPPs slow down upon collisions and accumulate in the dense phase. In our model, such a slowdown is unlikely to occur, and the SPPs therefore do not accumulate.

Related phase-separation behavior has also been reported in other systems consisting of particles with different motility in which MIPS is suppressed. In particular, the phase separation observed in the present system resembles those in underdamped mixtures of fast and slow SPPs with fixed propulsion speeds \cite{UnderdampFastAndSlowSPPs} and in overdamped mixtures of particles with different diffusivities \cite{DifferentDiffusivitiesDemix}. Indeed, MIPS is unlikely to occur in either of these systems, although for different reasons. In the former, inertia prevents particles from slowing down locally \cite{EffectOfInertiaOnMIPS}, while their speeds remain fixed and cannot be reduced by collisions. In the latter, by contrast, particles lack persistent motion.

A more detailed characterization of the ordering observed in the passive particle-rich phase remains an important subject for further study. In particular, it is necessary to quantify the degree of ordering and clarify how it depends on the self-propulsion speed. We are currently investigating these questions by analyzing the structural properties of the passive particle-rich phase. Such analyses will clarify the nature of the ordered state and help determine whether the observed ordering can be understood solely in terms of the speed-dependent effective excluded area or whether additional non-equilibrium effects contribute to its formation.

\section*{Acknowledgements}
This work was supported by the Japan Science and Technology Agency (JST) SPRING (Grant No. JPMJSP2136), the Japan Society for the Promotion of Science (JSPS) KAKENHI (Grant Nos. JP21K18604, JP23K04523, and JP22K03558), the National Science and Technology Council, Taiwan (Grant No. NSTC113-2112-M-001-426051-MY3), and the Taiwan International Graduate Program-X (TIGP-X) Pilot Program at Academia Sinica, Taiwan. The computations were performed using the Research Center for Computational Science, Okazaki, Japan (Project: 26-IMS-C078 and 25-IMS-C084) and the Research Institute for Information Technology, Kyushu University.
\clearpage

\appendix
\section{SINGLE–COMPONENT SPPs SYSTEM}\label{Appendix:single component SPPs system}

\begin{figure*}[t]
\centering
\includegraphics[width=\textwidth,clip]{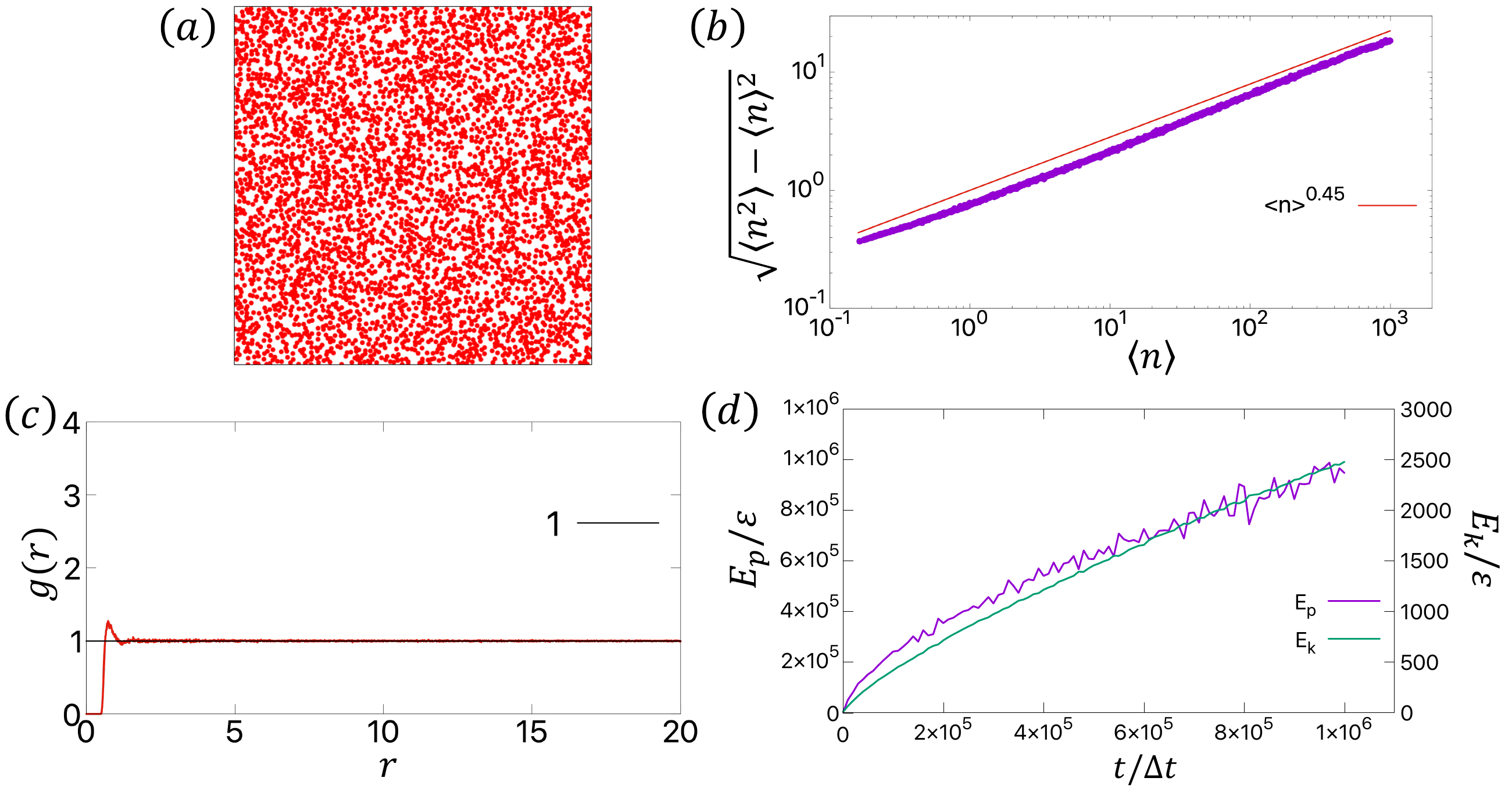}
\caption[]{The properties of a pure SPP system at ($v_0^{\mathrm{SP}}=3.0v_T$). (a) Snapshot of the system. (b) Number fluctuations $\sqrt{\langle n^2\rangle-\langle n\rangle^2}$, as a function of the mean number of particles $\langle n\rangle$. (c) Radial distribution function $g\left(r\right)$. (d) Time evolution of the total potential energy (left) and kinetic energy (right) , $E_p=\sum_{ij}V_{ij}$ and $E_k\equiv\sum_{i=1}^{N'_S}m\bm{v}_i^2/2{N_S}$.}
\label{Fig:only_SPP}    
\end{figure*}

To examine the behavior of a single–component system of SPPs, we performed a simulation with the same total number of particles and number density as those in the binary system, i.e., $N'_S=N_S+N_P$ and $N'_S/L'^2=0.700\sigma^{-2}$. We observed neither swarming nor MIPS in this system [Fig.~\ref{Fig:only_SPP}(a)]. Here, we show the result for the case of $v_0^{\mathrm{SP}}=3.0v_T$, which is the highest self–propulsion speed considered in this study.

The pure SPPs system exhibits a disordered structure. Figure~\ref{Fig:only_SPP}(b) shows the scaling behavior of number fluctuations. The standard deviation of the number of particles within a given region, $\sqrt{\langle n^2 \rangle-\langle n \rangle^2}$, increases as a power of the mean number of particles, $\langle n \rangle$. The scaling exponent is smaller than $1/2$, as predicted by the central limit theorem, indicating a uniform structure. Figure~\ref{Fig:only_SPP}(c) shows the radial distribution function. Its first peak is lower than that of the binary systems [Fig.~\ref{Fig:FirstPeakRDF}(c)], and the oscillations decay rapidly.

We should note that the total potential energy and kinetic energy ($E_p\equiv\sum_{ij}{V_{ij}}$ and $E_k\equiv\sum_{i=1}^{N'_S}m\bm{v}_i^2/2{N_S}$) do not reach a steady value because this system is not coupled to a thermostat [Fig.~\ref{Fig:FirstPeakRDF}(d)]. Consequently, it eventually becomes difficult to continue the simulation, and this point is different from that of the present SPP–passive particle mixtures. In this simulation, we set the time step to $\Delta t'=0.01\Delta t$ and simulated up to $10^6\Delta t$. For $t>10^5\Delta t$, $E_p$ and $E_k$ increase linearly with time, suggesting that the energy injected through self-propulsion is converted into closer approaches between particles within the repulsive interaction range. However, in this time region, we observed no significant changes in the particle structure. Thus, it seems reasonable to regard the system as being in a kind of disordered non-equilibrium steady state. We calculated the number fluctuations and the radial distribution function from snapshots taken after $t=5\times10^5 \Delta t$.

\section{COMPARING BINARY MIXTURES OF LARGE AND SMALL WCA PARTICLES}\label{Appendix:binary mixtures of large and small WCA-particles}
To compare the present phase separation with the phase separation of a mixture system consisting of large and small particles, we carried out additional simulations. We replaced the SPPs with larger passive particles. In this equilibrium system, all particles interact through the shifted WCA potential $V_{ij}\left(r\right)=4\epsilon\left[\left(\frac{\sigma}{r-\sigma_{\mathrm{ij}}+\sigma}\right)^{12}-\left(\frac{\sigma}{r-\sigma_{\mathrm{ij}}+\sigma}\right)^{6}\right]+\epsilon$, where $\sigma_{\mathrm{ij}}$ is the average diameter of $i$-th and $j$-th particles. The diameter of the replaced passive particles, $\sigma_{\mathrm{rep}}$, is varied. For clarity, we refer to the original passive particles as “original” particles and the inserted passive particles as “replaced” particles.

\begin{figure}[htbp]
\centering
\includegraphics[width=246pt,clip]{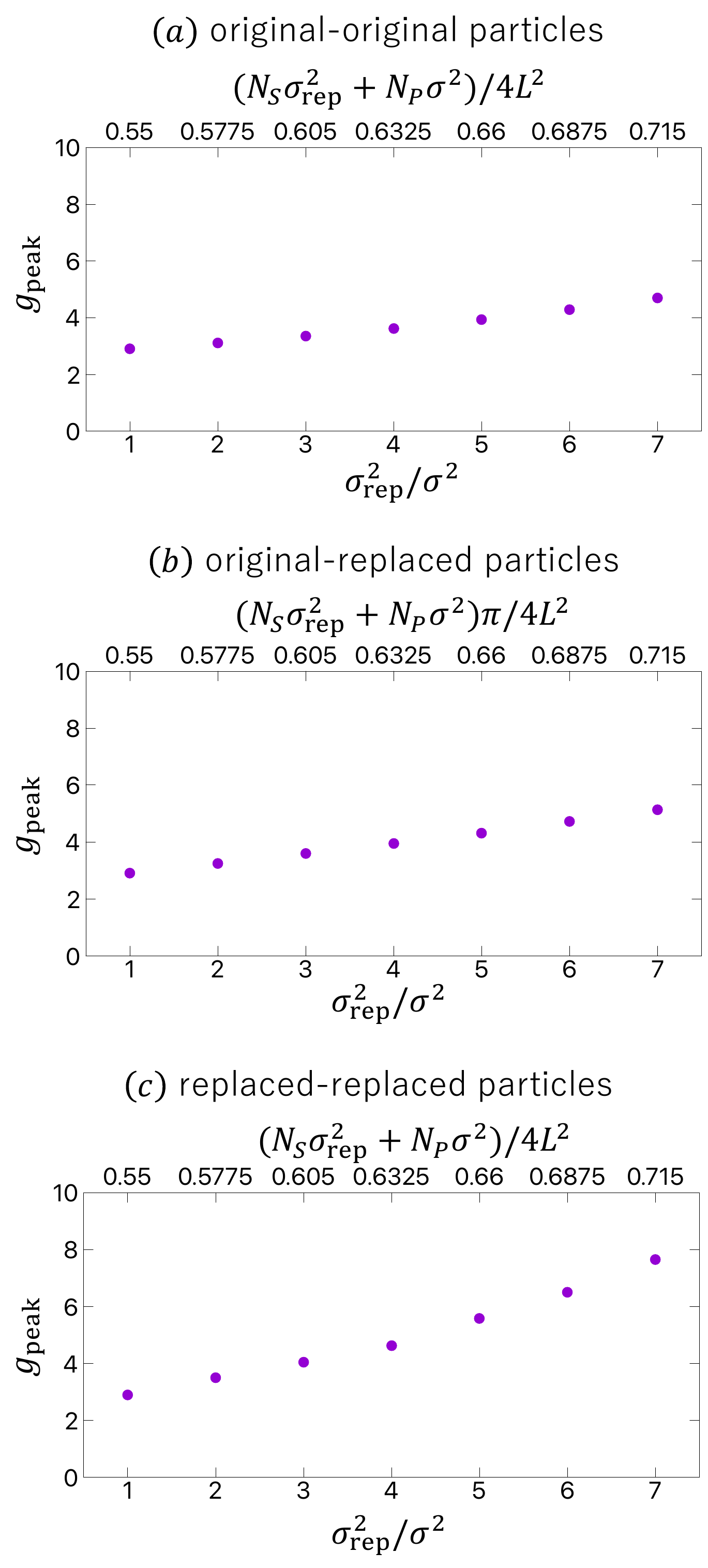}
\caption[]{The height of the first peaks of the radial distribution function (RDF) $g_{\mathrm{peak}}$. (a) $g_{\mathrm{peak}}$ between original particles (small), (b) $g_{\mathrm{peak}}$ between original and replaced particles, and (c) $g_{\mathrm{peak}}$ between replaced particles (large).}
\label{Fig:FirstPeakRDF_diluteLarge}
\end{figure}

We evaluated the height of the first peak of the RDF, $g_{\mathrm{peak}}$, for pairs of original–original, original–replaced, and replaced–replaced particles [Fig.~\ref{Fig:FirstPeakRDF_diluteLarge}]. The horizontal axis at the bottom represents the area of the replaced particles $\pi\sigma_{\mathrm{rep}}^2/4$ scaled by that of the original particles $\pi\sigma^2/4$. For reference, the horizontal axis at the top represents the packing fraction $\left(N_S\sigma_{\mathrm{rep}}^2+N_P\sigma^2\right)\pi/4L^2$.

In the additional simulations, the binary mixture of large and small particles exhibits different behavior. The $\sigma_{\mathrm{rep}}^2$ dependence of $g_{\mathrm{peak}}$ is shown in Fig.~\ref{Fig:FirstPeakRDF_diluteLarge}. For original–original (small–small) pairs [Fig.~\ref{Fig:FirstPeakRDF_diluteLarge}(a)], $g_{\mathrm{peak}}$ increases gradually as the size of the replaced (large) particles increases, in contrast to the abrupt increase observed in the active system [Fig.~\ref{Fig:FirstPeakRDF}(a)]. Similarly, $g_{\mathrm{peak}}$ between original and replaced particles [Fig.~\ref{Fig:FirstPeakRDF_diluteLarge}(b)] exhibits a gradual increase. This behavior is qualitatively different from that in the present active system [Fig.~\ref{Fig:FirstPeakRDF}(b)]. In this equilibrium system, the increase of $g_{\mathrm{peak}}$ between replaced particles is most significant [Fig.~\ref{Fig:FirstPeakRDF_diluteLarge}(c)], although the $g_{\mathrm{peak}}$ does not depend on the $v_0^{\mathrm{SP}}$ in the present active system [Fig.~\ref{Fig:FirstPeakRDF}(c)].

Here, we discuss the behavior of a binary mixture of SPPs and passive particles by comparing it with that of an equilibrium binary mixture of large and small particles. The SPPs do not form an ordered structure in the active system because their effective excluded regions can overlap as discussed in Sec.~\ref{Subsec:the radial distribution function}. In contrast, large WCA particles can scarcely approach one another within the contact distance, and the effective attraction between large particles is stronger than that between other particle pairs. Therefore, large particles form an ordered structure in equilibrium binary mixtures.

We should mention that in the equilibrium binary mixture examined here, the large particles did not show ordering because the size ratio and total packing fraction were insufficient. On the other hand, we can confirm that the effective attraction between large particles was stronger than that between large and small particles or between small particles. Furthermore, previous studies have demonstrated that large particles form an ordered structure when the radius ratio and the total packing fraction are sufficiently large \cite{36, 2DHDPS2}. Nevertheless, we could not perform calculations at such high ratios in this study because of the high computational cost.

Returning to the present active–passive mixture, passive particles exhibited ordering because the effective excluded volume generated by the SPPs was extremely large, and the passive particles could not approach one another more closely than the contact distance. A part of the passive particles approached the SPPs. We can confirm this behavior even when $v_0^{SP} = 3.0v_T$. However, the exclusion of the passive particles by the SPPs was strong. Indeed, the SPP–passive particle RDF [Fig.~\ref{Fig:RDF}(b)] showed that passive particles were excluded from the range of $1<r<20$. This exclusion confined the passive particles to a very limited space, resulting in a high packing fraction. Therefore, although the passive particles are effectively smaller than the SPPs, they formed an ordered structure because the repulsive-core interactions between the passive particles were sufficiently strong.

\section{THE AVERAGE KINETIC ENERGY OF SPPS}\label{Appendix:The average kinetic energy of SPPs}
\begin{figure}[htbp]
\centering
\includegraphics[width=246pt,clip]{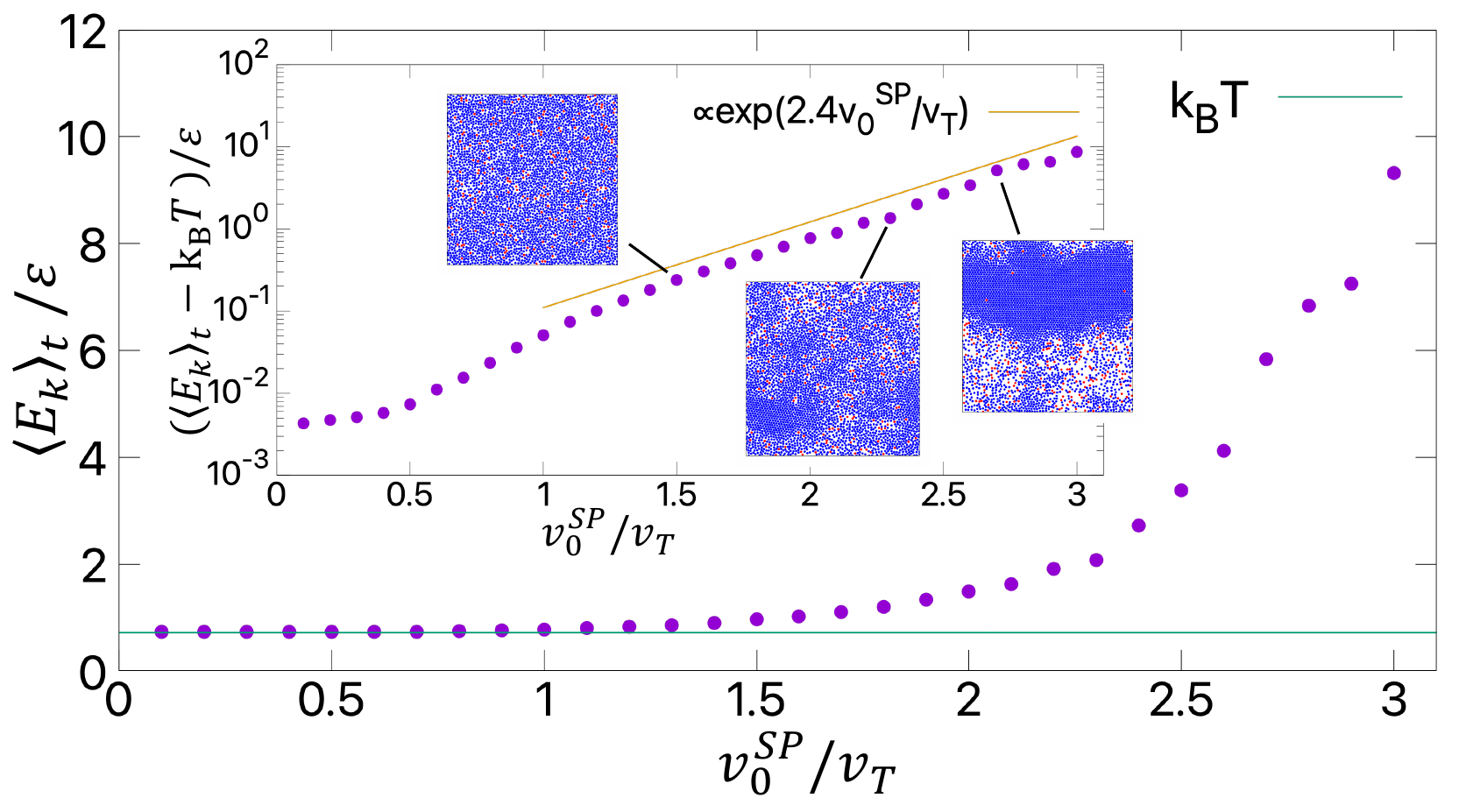}
\caption[]{Average kinetic energy of SPPs after $t=10^7$. The green line represents the target energy of thermostat, $\mathrm{k_B}T$. The inset shows the difference between $\left<E_k\right>_t$ and $\mathrm{k_B}T$ of the thermostat on a semi-log plot. The orange line represents the function of $\propto \exp{(2.4v_0^{\mathrm{SP}}/v_T)}$.}
\label{Fig:KineticEnegy_SPP_averaged_after1E+7}
\end{figure}

Surprisingly, the difference between the kinetic energy of SPP and $\mathrm{k_B}T$ increases exponentially with respect to the self-propulsion speed when $v_0^{\mathrm{SP}}\ge1.0v_T$. We obtained the average kinetic energy $\left<E_k\right>_t$ of SPPs in the steady state ($t>10^7 \Delta t$) and show $\left<E_k\right>_t$ as a function of $v_0^{\mathrm{SP}}$ in Fig.~\ref{Fig:KineticEnegy_SPP_averaged_after1E+7}. When the self-propulsion speed is slower than $1.0v_T$, $\left<E_k\right>_t$ is nearly equal to the thermostat’s target energy, $\mathrm{k_B}T$. In other words, when self-propulsion speed is low, we cannot distinguish SPPs from the passive particles in terms of the kinetic energy. On the other hand, the plot in Fig.~\ref{Fig:KineticEnegy_SPP_averaged_after1E+7} is bent between $v_0^{\mathrm{SP}}=0.5v_T$ and $v_0^{\mathrm{SP}}=1.0v_T$. When the self-propulsion speed exceeds $1.0v_T$, the difference between $\left<E_k\right>_t$ and $\mathrm{k_B}T$ increases exponentially [see inset in Fig.~\ref{Fig:KineticEnegy_SPP_averaged_after1E+7}]. We emphasize that this exponential growth is observed over a broad range of $v_0^{\mathrm{SP}}$, from the disorder regime to the phase-separated regime.

As discussed in the preceding paragraph, at high self-propulsion speeds, the kinetic energy is not proportional to the square of the speed, as would be expected from the conventional relationship between kinetic energy and velocity. The pushing behavior of SPPs provides a qualitative explanation for this phenomenon. As the self-propulsion speed increases, SPPs push surrounding particles away more strongly, thereby expanding the region in which they can move freely. The SPP is strongly pushed back by the passive particles surrounding the caged area. As a result, the net velocity of SPPs increases beyond that expected from an increase in $v_0^{\mathrm{SP}}$. However, this mechanism is not yet fully understood, especially the exponential increase. Further analysis is required to clarify the connection.

\clearpage
\phantomsection
\addcontentsline{toc}{section}{References} 
\renewcommand{\refname}{References}        

\end{document}